\documentclass{epl}

\title{Generalized synchronization onset\footnote{This
  paper was published in EPL. 2005. V. 72, No 6.
  p. 901--907}}
\shorttitle{GS onset}
\author{A.E. Hramov, A.A. Koronovskii \and O.I. Moskalenko}
\institute{Faculty of Nonlinear Processes, Saratov State
University, Astrakhanskaya, 83, Saratov, 410012, Russia}



\usepackage{amssymb}
\begin{document}

\maketitle

\begin{abstract}
The behavior of two unidirectionally coupled chaotic oscillators
near the generalized synchronization onset has been considered.
The character of the boundaries of the generalized synchronization
regime has been explained by means of the modified system
approach.
\end{abstract}

Chaotic synchronization is one of the fundamental nonlinear
phenomena actively studied
recently~\cite{Pikovsky:2002_SynhroBook}. Several different types
of chaotic synchronization of coupled oscillators, i.e.
\emph{generalized synchronization} (GS)
\cite{Rulkov:1995_GeneralSynchro}, \emph{phase synchronization}
(PS) \cite{Pikovsky:2002_SynhroBook}, \emph{lag synchronization}
(LS) \cite{Rosenblum:1997_LagSynchro}, \emph{complete
synchronization} (CS) \cite{Pecora:1990_ChaosSynchro}  and
\emph{time scale synchronization} (TSS)
\cite{Hramov:2005_TSSPhysD} are well known. There are also
attempts to find unifying framework for chaotic synchronization
of coupled dynamical systems \cite{Brown:2000_ChaosSynchro,%
Boccaletti:2002_SynchroPhysReport,Boccaletti:2001_UnifingSynchro,%
Hramov:2004_Chaos}.

One of the interesting and intricate types of the synchronous
behavior of unidirectionally coupled chaotic oscillators is the
generalized synchronization. The presence of GS between the
response $\mathbf{x}_{r}(t)$ and drive $\mathbf{x}_{d}(t)$ chaotic
systems means that there is some functional relation
${\mathbf{x}_r(t)=\mathbf{F}[\mathbf{x}_d(t)]}$ between system
states after the transient finished. This functional relation
$\mathbf{F}[\cdot]$ may be smooth or fractal. According to the
properties of this relation, GS may be divided into the strong
synchronization and weak synchronization,
respectively~\cite{Pyragas:1996_WeakAndStrongSynchro}. There are
several methods to detect the presence of GS between chaotic
oscillators, such as the auxiliary system
approach~\cite{Rulkov:1996_AuxiliarySystem} or the method of
nearest neighbors~\cite{Rulkov:1995_GeneralSynchro,%
Pecora:1995_statistics}. It is also possible to calculate the
\emph{conditional Lyapunov exponents} (CLEs)~\cite{Pecora:1991_ChaosSynchro,%
Pyragas:1996_WeakAndStrongSynchro} to detect GS. The regimes of LS
and CS are also the particular cases of GS.

One of the interesting aspects of the generalized synchronization
study is the analysis of the onset of this regime. In particular,
the intermittent behavior is known to be revealed on the onset of
GS~\cite{Hramov:2005_IGS_EuroPhysicsLetters,Zhan:2003_IGS} as well
as in case of the lag~\cite{Rosenblum:1997_LagSynchro,%
Boccaletti:2000_IntermitLagSynchro,Zhan:2002_ILS}
or phase~\cite{Pikovsky:1997_EyeletIntermitt,%
Pikovsky:1997_PhaseSynchro_UPOs,Lee:1998:PhaseJumps}
synchronization.

At the same time, there are known examples of the unidirectionally
coupled chaotic systems for which the location of the generalized
synchronization onset on the ``parameter mismatch
--- coupling strength'' plane differs radically from the other
synchronization types. Indeed, for two unidirectionally coupled
R\"ossler oscillators with identical control parameters the value
of the coupling strength corresponding to the onset of GS is twice
as much as for the same oscillators with parameters detuned
sufficiently~\cite{Zhigang:2000_GSversusPS}.  For two
unidirectionally coupled one--dimensional complex Ginzburg--Landau
equations with sufficiently detuned control parameters the
threshold of the GS regime onset does not depend on the value of
the drive system parameter when the response system parameters are
fixed~\cite{Hramov:2005_GLEsPRE}. Alternatively, for all other
types of chaotic synchronization the dependence of the threshold
of the synchronous regime arising on the value of the control
parameter mismatch behaves in the different way, i.e. when the
control parameter mismatch decreases the coupling parameter value
corresponding to the onset of the synchronous regime also
declines.

This letter aims to explain the mechanisms resulting in the
generalized synchronization arising in the unidirectionally
coupled chaotic oscillators and determining the location of the GS
onset.

The causes of the generalized synchronization arising may be
clarified by means of a modified system approach proposed
in~\cite{Hramov:2005_GSNature}. Let us consider the behavior of
two unidirectionally coupled chaotic oscillators with slightly
mismatched parameters
\begin{equation}
\begin{array}{l}
\mathbf{\dot x}_d=\mathbf{H}(\mathbf{x}_d,\mathbf{g}_d)\\
\mathbf{\dot x}_r=\mathbf{H}(\mathbf{x}_r,\mathbf{g}_r)+
\varepsilon\mathbf{A}(\mathbf{x}_d-\mathbf{x}_r),
\end{array}
\label{eq:Oscillators1}
\end{equation}
where $\mathbf{x}_{d,r}$ are the state vectors of the drive and
response systems, respectively; $\mathbf{H}$ defines the vector
field of the considered systems, $\mathbf{g}_d$ and $\mathbf{g}_r$
are parameters vectors, $\mathbf{A}={\{\delta_{ij}\}}$ is a
coupling matrix ($\delta_{ii}=0$ or $1$, $\delta_{ij}=0$ ($i\neq
j$)), $\varepsilon$ is a scale parameter characterizing the
coupling strength.

In this case one can see that the response system
$\mathbf{x}_r(t)$ may be considered as a modified system
\begin{equation}
\mathbf{\dot x}_m(t)=\mathbf{H}'(\mathbf{x}_m(t), \varepsilon)
\label{eq:RsOsc}
\end{equation}
(where $\mathbf{H}'(\mathbf{x}(t))=
\mathbf{H}(\mathbf{x}(t))-\varepsilon\mathbf{A}\mathbf{x}(t)$)
under the external force $\varepsilon\mathbf{A}\mathbf{x}_d(t)$:
\begin{equation}
\mathbf{\dot u}_m(t)=\mathbf{H}'(\mathbf{u}_m(t),\varepsilon)+
\varepsilon\mathbf{A}\mathbf{x}_d(t), \label{eq:RsOsc&Force}
\end{equation}
It is easy to see that the term
$-\varepsilon\mathbf{A}\mathbf{x}(t)$ brings the additional
dissipation into the system~(\ref{eq:RsOsc}). Indeed, the phase
flow contraction is characterized by means of the vector field
divergence. Obviously, the vector field divergences of the
modified and the response systems are related with each other as
\begin{equation}
\mathrm{div}\,\mathbf{H}'=\mathrm{div}\,\mathbf{H}
-\varepsilon\sum\limits_{i=1}^N\delta_{ii}
\end{equation}
(where $N$ is the dimension of the modified system phase space),
respectively. So, the dissipation in the modified system is
greater than in the response one and it increases with the growth
of the coupling strength $\varepsilon$.

The generalized synchronization regime arising
in~(\ref{eq:Oscillators1}) may be considered as a result of two
cooperative processes taking place simultaneously. The first of
them is the growth of the dissipation in the
system~(\ref{eq:RsOsc}) and the second one is an increase of the
amplitude of the external signal. Evidently, both processes are
correlated with each other by means of parameter $\varepsilon$ and
can not be realized in the coupled oscillator
system~(\ref{eq:Oscillators1}) independently. Nevertheless, it is
clear, that an increase of the dissipation in the modified
system~(\ref{eq:RsOsc}) results in the simplification of its
behavior and the transition from the chaotic oscillations to the
periodic ones. Moreover, if the additional dissipation is large
enough the stationary fixed state may be realized in the modified
system. On the contrary, the external chaotic force
$\varepsilon\mathbf{A}\mathbf{x}(t)$ tends to complicate the
behavior of the modified system and impose its own dynamics on it.
Obviously, the generalized synchronization regime may not appear
unless own chaotic dynamics of the modified system is suppressed.

Note also, that the onset of GS is determined by the stability of
the periodic regimes of the modified system~(\ref{eq:RsOsc}) which
does not seem to depend on the mismatch of the control parameters
$\mathbf{g}_{d,r}$ of the unidirectionally coupled oscillators.
The stability of the periodical regimes is caused by the property
of the modified system only. Therefore, the value of
$\varepsilon_{GS}$ is not supposed to depend greatly on the
parameter mistuning\footnote{This conclusion agrees well with
numerical results of \cite{Hramov:2005_GSNature}.}. Nevertheless,
the sensitive dependence of the coupling strength value
$\varepsilon_{GS}$ corresponding to the onset of GS on the control
parameter mismatch may take place as it has been discussed above
(see also~\cite{Zhigang:2000_GSversusPS}).

To explain the mechanisms causing such dependence of the coupling
strength on the control parameter mismatch let us consider two
unidirectionally coupled R\"ossler oscillators
\begin{equation}
\begin{array}{ll}
\dot x_{d}=-\omega_{d}y_{d}-z_{d},& \dot x_{r}=-\omega_{r}y_{r}-z_{r} +\varepsilon(x_{d}-x_{r}),\\
\dot y_{d}=\omega_{d}x_{d}+ay_{d},& \dot y_{r}=\omega_{r}x_{r}+ay_{r},\\
\dot z_{d}=p+z_{d}(x_{d}-c),\mbox{\qquad\qquad\qquad} & \dot z_{r}=p+z_{r}(x_{r}-c),\\
\end{array}
\label{eq:Roesslers}
\end{equation}
where $\varepsilon$ is a coupling parameter. The control parameter
values have been selected by analogy
with~\cite{Zhigang:2000_GSversusPS} as $a=0.15$, $p=0.2$,
$c=10.0$. The $\omega_r$--parameter determining the main frequency
of the response system has been selected as $\omega_r=0.95$, and
the analogous parameter $\omega_d$ of the drive system has been
varied in the range from 0.8 to 1.1 providing the slight mismatch
of the interacting oscillators.

\begin{figure}[tb]
\centerline{\includegraphics*[scale=0.5]{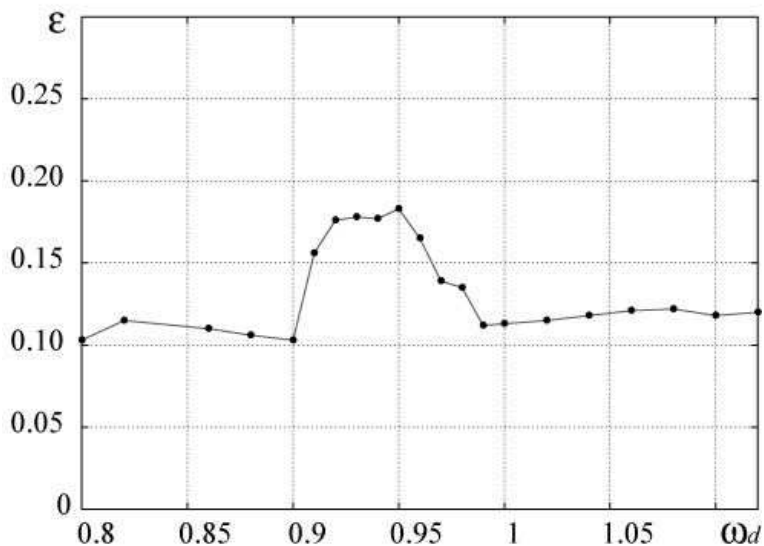}} \caption{The
boundary of the GS onset on the $(\omega_d,\varepsilon)$ parameter
plane for two unidirectionally coupled R\"ossler
systems~(\ref{eq:Roesslers}). The $\omega$--parameter of the
response system has been fixed as $\omega_r=0.95$.
\label{fgr:GSBoundary}}
\end{figure}

The boundary of the GS regime in system~(\ref{eq:Roesslers}) on
the $(\omega_d,\varepsilon)$--plane is shown in
Fig.~\ref{fgr:GSBoundary}. The GS onset has been detected by the
conditional Lyapunov exponent
computation~\cite{Pyragas:1996_WeakAndStrongSynchro} and verified
by the auxiliary system method~\cite{Rulkov:1996_AuxiliarySystem}.
The threshold of the GS arising appears to be essentially higher
for the small control parameter detuning of the considered systems
rather than for the large ones. At the same time, if the parameter
mismatch $(\omega_d-\omega_r)$ is large enough, the coupling
strength value $\varepsilon_{GS}$ corresponding to the GS regime
onset is almost independent on the value $\omega_d$ of the drive
system (see Fig.~\ref{fgr:GSBoundary}).

Such system behavior may be explained by means of the modified
system method discussed in detail above. The response system
(\ref{eq:Roesslers}) can  be reduced to the modified system
\begin{equation}
\begin{array}{l}
\dot x_{m}=-\omega_{r}y_{m}-z_{m} -\varepsilon^* x_{m},\\
\dot y_{m}=\omega_{r}x_{m}+ay_{m},\\
\dot z_{m}=p+z_{m}(x_{m}-c).
\end{array}
\label{eq:ModRoesslers}
\end{equation}
To separate the discussed above different mechanisms resulting in
the GS arising the $\varepsilon^*$ parameter has been used instead
of $\varepsilon$.

In Fig~2 the bifurcation diagram for the modified
system~(\ref{eq:ModRoesslers}) is shown. It is clear that with the
increase of the $\varepsilon^*$ parameter this system undergoes
the transition from the chaotic to periodic oscillations through
the inverse cascade of the period doubling. One can easily see,
that starting from the value $\varepsilon^*_p\approx 0.06$ the
periodic oscillations take place in the modified
system~(\ref{eq:ModRoesslers}). So, if the $\varepsilon^*$
parameter value is large enough the autonomous modified system
displays the periodic oscillations.

To study the GS arising the non-autonomous dynamics of the
modified system~(\ref{eq:ModRoesslers})
\begin{equation}
\begin{array}{l}
\dot x_{m}=-\omega_{r}y_{m}-z_{m} -\varepsilon^* x_{m}+\varepsilon F(t),\\
\dot y_{m}=\omega_{r}x_{m}+ay_{m},\\
\dot z_{m}=p+z_{m}(x_{m}-c),
\end{array}
\label{eq:ForcedModRoesslers}
\end{equation}
should be considered, the time series $x_d(t)$ of the drive system
being used as the external force $F(t)$.

For the control parameters pointed above the Fourier spectrum of
the drive system~(\ref{eq:Roesslers}) is characterized by the
single well--defined frequency component $f_m$. When varying the
$\omega_d$-parameter the main frequency $f_m$ also changes.
Therefore, the behavior of the modified system
(\ref{eq:ForcedModRoesslers}) under the external harmonic signal
$F(t)=A\cos(2\pi f_mt)$ may be considered as the first
approximation of the dynamics of two unidirectionally coupled
chaotic oscillators~(\ref{eq:Roesslers}). The amplitude $A$ of the
external harmonic signal $F(t)$ should correspond to the drive
R\"ossler system behavior.

\begin{figure}[tb]
\centerline{\includegraphics*[scale=0.5]{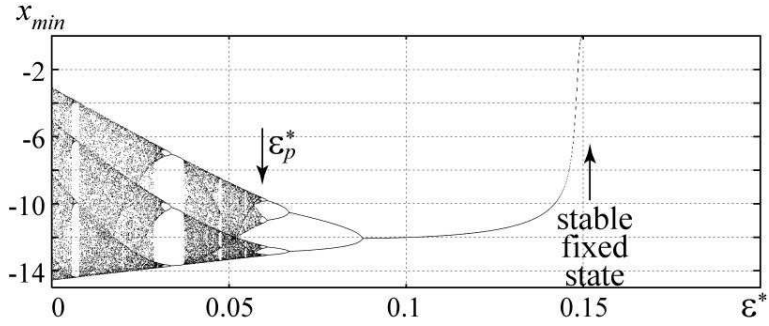}} \caption{The
bifurcation diagram of the modified R\"ossler
system~(\ref{eq:ModRoesslers}) versus the parameter
$\varepsilon^*$. The value of the parameter $\varepsilon^*$ when
the modified system starts demonstrating the periodic dynamics is
shown by an arrow. \label{fgr:BDDiagram}}
\end{figure}

Obviously, when the main frequency of the response system and the
frequency $f_m$ of the external signal are close enough with each
other the frequency entrainment may take place resulting in the
synchronization. In this case the behavior of the modified system
(and the response one, respectively) may be different
qualitatively inside the synchronization tongue and outside it.
The location of the boundaries of the GS regime (see
Fig.~\ref{fgr:GSBoundary}) confirms this assumption. When the
parameter mismatch of the considered response and drive
oscillators is large enough (the main frequencies are not
entrained and the system~(\ref{eq:ForcedModRoesslers}) is outside
of the synchronization tongue) the GS regime onset is observed
practically for the same value of the coupling strength
$\varepsilon$ independently of the $\omega_d$ parameter value. In
this case the threshold of the GS regime arising is defined mainly
by the properties of the modified
system~(\ref{eq:ForcedModRoesslers}). Obviously, the features of
the GS onset location for the small mistuning of the control
parameters may be defined by the system dynamics inside the
synchronization tongue.

Let us consider the behavior of the non-autonomous modified
system~(\ref{eq:ForcedModRoesslers}) under the external harmonic
signal in detail. The value of the $\varepsilon^*$ parameter has
been fixed as $\varepsilon^*=0.11$ that corresponds approximately
to the threshold of the GS arising when the control parameters of
the coupled R\"ossler oscillators are detuned sufficiently. The
amplitude of the external harmonic signal has been used in the
form $A=\varepsilon B$, the value of $B$ parameter has been
selected as $B\approx 0.105$ for the levels of the energy of the
harmonic signal $F(t)=A\cos(2\pi f_mt)$ and the main spectral
component of the Fourier spectrum of the drive system
$\mathbf{x}_d(t)$ to be equal. In this case the $\varepsilon$
parameter determines the intensity of the external influence on
the modified system and may be considered as a quantity
corresponding to the coupling strength of the coupled oscillators
(\ref{eq:Roesslers}). Such normalization allows us to mark the
synchronization region of the non-autonomous modified system on
the $(\omega_d, \varepsilon)$ directly.

\begin{figure}[tb]
\centerline{\includegraphics*[scale=0.5]{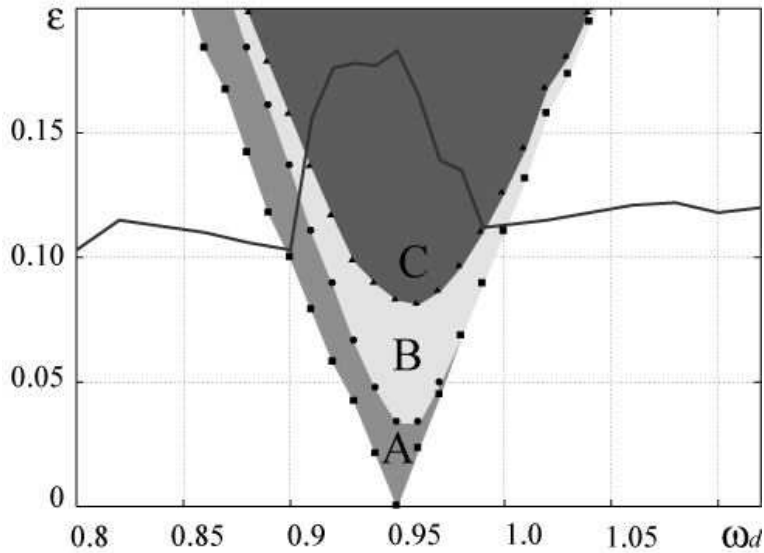}} \caption{The
synchronization tongue of the non-autonomous modified system under
the external harmonic signal on the $(\omega_d,\varepsilon)$
plane. The area of the period--2 regime (marked by the letter
``B'') as well as the region of the chaotic oscillations (marked
by ``C'') inside the synchronization tongue are shown by the
different colors. The boundary of the GS onset (compare with
Fig.~\ref{fgr:GSBoundary}) has also been shown.
\label{fgr:SynchroTongue}}
\end{figure}

The synchronization tongue of the non-autonomous modified
system~(\ref{eq:ForcedModRoesslers}) on the
$(\omega_d;\varepsilon)$ plane is shown in
Fig.~\ref{fgr:SynchroTongue}. The region \textrm{A} presents the
set of the parameter values corresponding to the periodic
oscillation regime with the frequency $f_m$ of the external
harmonic signal. With the increase of the $\varepsilon$ parameter
value inside this region the system undergoes the transition to
chaotic behavior through the cascade of the period doubling. The
region \textrm{C} corresponds to the chaotic oscillations, while
the area \textrm{B} presents the parameter values of the period
doubling cascade. The boundary of the GS regime is also shown in
Fig.~\ref{fgr:SynchroTongue} for the cause of the dependence of
the GS arising threshold $\varepsilon_{GS}$ on the control
parameter mismatch to be explained clearly. One can see easily,
that the behavior of the response system differs essentially
inside and outside the synchronization tongue of the modified
system.

Let us consider briefly the cause of such location of the GS
onset. The external signal (chaotic time series of the drive
R\"ossler system or the simulating harmonic signal) excites the
proper chaotic dynamics of the modified system. Therefore, for the
considered coupled R\"ossler systems the GS regime starts being
destroyed with the growth (inside the synchronization tongue) of
the $\varepsilon$ parameter. To suppress the chaotic oscillations
in the modified system the value of the dissipation
$\varepsilon^*$ should be increased. As in the coupled
systems~(\ref{eq:Roesslers}) the coupling strength $\varepsilon$
plays simultaneously the role both of the dissipation and of the
intensity of the drive system signal, the suppression of the
proper chaotic dynamics of the response system takes place for the
more larger values of the parameter $\varepsilon$. Therefore, the
threshold of the GS regime arising is shifted towards the large
values of the coupling strength $\varepsilon$.

Comparing the boundary of the GS regime in the coupled R\"ossler
systems (\ref{eq:Roesslers}) with the location of the
synchronization tongue of the non-autonomous modified
system~(\ref{eq:ForcedModRoesslers}) under the external harmonic
signal on the $(\omega_d,\varepsilon)$ plane it is necessary to
take into account that the synchronization area has been obtained
for the fixed value of the parameter $\varepsilon^*$. Moreover,
the chaotic dynamics of the drive system influencing on the
location of the GS onset has also been eliminated from the
consideration. Therefore, this synchronization tongue on the
$(\omega_d,\varepsilon)$ plane has only the qualitative character
allowing to explain the mechanisms resulting in the GS onset
features.

In conclusion, we have explained the peculiarity of the GS onset
in the unidirectionally coupled R\"ossler oscillators by means of
the modified system dynamics consideration. The character of the
GS onset location is determined by the entrainment (or, on the
contrary, by the absence of it) of the main frequencies of the
Fourier spectra of the interacting systems. Inside the
synchronization tongue the GS onset may be shifted towards the
large values of the coupling strength if the external influence
excites the proper chaotic dynamics inside the area of the main
frequencies entrainment. It is important to note that this
conclusion has been made for the chaotic systems with a distinct
main frequency in the power spectrum, so, the discussed phenomena
and the presented arguments are specific to such systems. Indeed,
the mechanism discussed above is determined by the interaction of
the main spectral components of coupled chaotic oscillators.
Probably, one can expect that the phenomenon of the increase of
the coupling strength $\varepsilon_{GS}$ corresponding to the
onset of the generalized synchronization regime for the identical
parameter values in comparison with the detuned ones may not be
observed for the system with broad power spectra without distinct
main frequency.

\acknowledgments

We thank Svetlana V. Eremina for the English language  support. We
thank the referees for providing very helpful comments and
advices. This work has been supported by U.S.~Civilian Research \&
Development Foundation for the Independent States of the Former
Soviet Union (CRDF, grant {REC--006}), Russian Foundation of Basic
Research (project 05--02--16273). We also thank ``Dynastiya''
Foundation.

\end{document}